\documentclass[nofootinbib]{appolb}
\usepackage{amssymb}  \usepackage{color}
\usepackage{graphicx}               \usepackage{amsfonts}
\usepackage{txfonts}                  \usepackage{times}
\usepackage{subfigure}              \usepackage{geometry}
\newcommand{\beq}{\begin{equation}}     \newcommand{\eeq}{\end{equation}}
\newcommand{\beqa}{\begin{eqnarray}}    \newcommand{\eeqa}{\end{eqnarray}}
\newcommand{\bde}{\begin{description}}  \newcommand{\ede}{\end{description}}
\newcommand{\ben}{\begin{enumerate}}    \newcommand{\een}{\end{enumerate}}

\newcommand{\la}{\langle}               \newcommand{\ra}{\rangle}
\newcommand{\kT}{{k_{\rm B}T} } 
\newcommand{\bm}[1]{\mbox{\boldmath ${#1}$}}

\newcommand{\eqn}[1]{\beq{ #1 }\eeq}
\newcommand{\inv}[1]{{\frac{1}{#1}}}
\newcommand{\inRbracket}[1]{{\left({#1}\right)}}
\newcommand{\inSbracket}[1]{{\left[{#1}\right]}}

\begin{document}
\title{From adiabatic piston to non-equilibrium hydrodynamics}
\thanks{Presented at 25th Smoluchowski symposium, September, 2012 in Krakow.%
}
\author{{Ken Sekimoto$^{1,2}$, Antoine Fruleux$^{2,1}$, Ryoichi Kawai$^3$ 
and Nathan Ridling$^3$}
\address{$^1$Mati\`eres et Syst\`emes Complexes, CNRS-UMR7057,
Universit\'e Paris-Diderot, 75205 Paris, France,
\\$^2$Gulliver, CNRS-UMR7083, ESPCI, 75231 Paris, France,\\$^3$Department of Physics, University of Alabama at Birmingham, }
}
\maketitle
\begin{abstract}
Based on the new concept of the {\it momentum transfer deficiency due to dissipation} (MDD),
the physical basis of the mechanism of ``adiabatic piston'' is explained.
The implication of MDD in terms of hydrodynamics under non-equilibrium steady state also discussed.
\end{abstract}

\section{Introduction}
Since the 1920's a simple question associated to non-equilibrium statistical physics has been addressed: If we put a Brownian piston of mass $M$ between the two semi-infinite cylinders each being filled with an ideal gas consisting of particles with mass $m(\ll M)$, what is the non-equilibrium steady state ?
Here the temperature and pressure of the gas in the left cylinder are prepared at $(T,p)$ while those in the right cylinder are at $(T',p)$. The surface area of the piston is the same on both sides.
It is clear that, if the piston were firmly fixed and if the piston is ``adiabatic'', then there would be no net force on the piston because the gas in each cylinder remains in equilibrium and presses the piston by the same pressure but in the opposing directions.
When the Brownian motion of this adiabatic piston is allowed, however, this motion will allow the energy transfer from the hotter gas (e.g. the side of $T$ if $T>T'$) to the colder gas ({\it ibid.} $T'$) \cite{feynman63}. The question is if this non-equilibrium precess leads to a non-vanishing net force on the piston.
The macroscopic thermodynamics cannot answer this type of question \cite{Callen1}, neither the Langevin description can give an answer to this type of setup \cite{VdB-prl2004-ratchet}. While the stochastic energetics \cite{sekimoto97,LNP} can describe correctly the heat flow, the non-equilibrium force is beyond the resolution of this level of description.

Many calculative studies have been reported in the past both on this problem and also on a class of Brownian ratchet models, which turned out to be essentially the same problem as adiabatic piston \cite{AFRKKS2012}. 
All these studies have been done using either by  {\it ad hoc} treatment of Master-Boltzmann equations with truncated moment hierarchical expansion with $\epsilon=\sqrt{m/M}$ as small parameter, or by molecular dynamics (MD) simulations, see the references cited in \cite{AFRKKS2012}. Both the perturbative approach and the MD simulation consistently concluded that the Brownian piston will move steadily towards the hotter gas. Nevertheless a clear physical understanding was missing. A frequently given hand-waving argument was that the hotter side losing the heat has locally lower pressure. But it is not a valid argument. If the ideal gas is  used, the cooled particle will never hit again the Brownian piston while the freshly colliding particles are characterized by the equilibrium parameters, $(T,p)$ or $(T',p)$. The adiabatic piston has, therefore, remained among { "Some problems in statistical mechanics that I would like to see solved"} \cite{Lieb1999}.

It is only very recently \cite{AFRKKS2012} that the physical explanation to the above question was definitely given. The key is to take into account the interplay between the energy transfer and the momentum transfer at the gas-piston interfaces.
Once this point is understood, the results of elaborated perturbative calculations could be perfectly  reproduced just by a few lines' calculations, except for an overall numerical factor.
The purpose of the present paper is to summary the basic idea and discuss its generalization.
The organization of the paper is the following: 
In the next section (\S~\ref{sec:mdd}) we recapitulate the main line of this mechanism. 
Especially the key concept of the momentum transfer deficit due to dissipation (MDD) is explained using a simple argument. The relation to the traditional calculative approach is also mentioned.
As a prologue to the extension to the dense gas case,
we introduce in \S~\ref{sec:toy} a toy model that shows how the energy flow and momentum flow having different symmetry in space and time can make the uniform pressure and the heat conduction compatible.
In \S~\ref{sec:hydro} we show the implication of the MDD to the non-equilibrium hydrodynamics of dense hard-core gas.

\section{Review of the physics of adiabatic piston\label{sec:mdd}}
We outline the concept of MDD as the underlining mechanism of the adiabatic piston. The readers might refer to \cite{AFRKKS2012} \cite{MDD-Kawai2012} for the technical details and its generalization to inelastic case.

\begin{figure}[h] \centerline{\includegraphics[width=4cm]{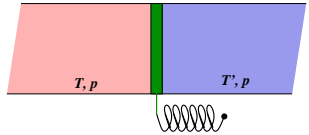}}
\caption{Trapped adiabatic piston.} 
\label{fig:trapped-piston.eps}
\end{figure}

The essential point of the adiabatic piston is more clearly grasped when the Brownian piston separating the ideal gases is trapped by a potential force such as an elastic spring (Fig.~\ref{fig:trapped-piston.eps}) so that the mean velocity of the piston vanishes ($\la V\ra=0$). If there appears a non-equilibrium force $F_{\rm NESS}$ on this trapped Brownian piston, the steady state velocity $\la V\ra$  of the piston in the absence of trapping is given by the balance with the passive frictional force, 
$F_{\rm NESS}-(\gamma+\gamma')\la V\ra=0,$ where $\gamma$ and $\gamma'$ are the 
friction coefficient of the Brownian piston against the respective gas.

The first step is to realize that the Brownian motion of the piston serves merely as the mediator of the energy transfer, or heat, from hotter gas to the cooler gas. While correlation between collisions with the piston by the hot gas particles and by the cold gas particles are essential for the irreversibility of this purely mechanical problem, we can bypass all the details for the purpose of understanding the non-equilibrium force $F_{\rm NESS}$.
The rate of the energy transfer per unit surface of the piston, $j^{(e)}$ can be found using the stochastic energetics \cite{LNP} or even by a heuristic argument \cite{parrondo96}. The result reads, 
$j^{(e)}=({k_{\rm B}T-k_{\rm B}T'})/[{M(\tilde{\gamma}^{-1}+\tilde{\gamma}^{\prime -1})}], $
where $\tilde{\gamma}=\gamma/A$ and $\tilde{\gamma}'=\gamma'/A$ with $A$ being the area of each piston surface.

\begin{figure}[h] \centerline{\includegraphics[width=4cm]{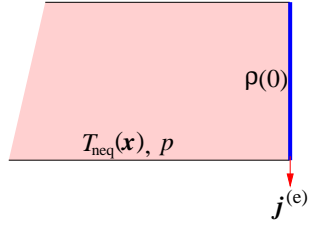}}
\caption{Gas in contact with an energy-transferring wall at $x=0$.
The contact density is denoted by $\rho(0)$. In general the effective temperature $T_{\rm new}(x)$ should depend on the position $x$.} 
\label{fig:neq-wall.eps}
\end{figure}

Once we know the energy transfer rate across the gas-piston interface,
we can concentrate on the following problem: 
When an ideal gas prepared in the equilibrium characterized by $(T,p)$ is 
brought into contact with the wall that absorbs (or injects) energy at the rate 
$j^{(e)}$ per unit surface (Fig.~\ref{fig:neq-wall.eps}), how the pressure on the wall is modified from $p$?
Consider that gas particles with a typical velocity component normal to the wall 
 $v^\perp_{\rm in}$ collides the energy-transferring wall. They are reflected back with a velocity
  $v^\perp_{\rm out}.$  While the typical incoming velocity should be the thermal velocity
$v^\perp_{\rm in}=v_{\rm th}=\sqrt{\kT/m},$ except for the numerical factor, 
the typical outgoing velocity depends on the energy transfer rate 
 $j^{(e)}$ and the collision rate $\nu_{\rm col}$ per unit area on the energy-transferring wall through the energy balance condition:
\[\frac{\bm{j}^{(e)}}{\nu_{\rm col}}=\frac{m}{2}{v^\perp_{\rm in}}^2-\frac{m}{2}{v^\perp_{\rm out}}^2 
=\frac{m}{2}{v_{\rm th}}^2-\frac{m}{2}{v^\perp_{\rm out}}^2. \]
Assuming that the energy transfer is sufficient small, the
 right hand side (r.h.s.) of the above equation is approximated as 
\[ (mv^\perp_{\rm in}-m|v^\perp_{\rm out}|)\frac{v^\perp_{\rm in}+|v^\perp_{\rm out}|}{2}\simeq(mv^\perp_{\rm in}-m|v^\perp_{\rm out}|){v_{\rm th}}.\]
Substituting this result into the above equation, we find 
\eqn{\label{eq:MDD}(mv^\perp_{\rm in}-m|v^\perp_{\rm out}|)\nu_{\rm col}\simeq \frac{\bm{j}^{(e)}}{v_{\rm th}} }
The left hand side (l.h.s.) of this relation gives the momentum transfer deficit due to dissipation (MDD). In other words, upon the collision, the gas particles kicks the wall less strongly in non-equilibrium than equilibrium if a part of their incoming kinetic energy was taken out by the energy-transferring wall.
In terms of the net momentum transfer rate per unit surface, ${\sf j}^{(p)\perp\,\perp}$
$=\nu_{\rm col}(mv_{\rm th}+|v^\perp_{\rm out}|),$ 
the equilibrium value, $p=2mv_{\rm th}\nu_{\rm col},$ is corrected by this MDD to give  
\[
{\sf j}^{(p)\perp\,\perp}=p-\frac{\bm{j}^{(e)}}{v_{\rm th}} .\]

In retrospect, the traditional approach through the Master-Boltzmann equation could have given the same insight. For the setup of Fig.~\ref{fig:neq-wall.eps}, this equation can be written as follows:
\beqa
&& \partial_t P(X,V,t)= -V\partial_X P(X,V,t)
-[-\gamma_bV-\partial_X U(X)]\partial_V P(X,V,t) 
\cr && \quad
-\int_{V'}W(V'|V)P(X,V,t)
+ \int_{V'} W(V|V')P(X,V',t)+\frac{\kT_b}{\gamma_b}\partial_X^2 P(X,V,t),
\nonumber
\eeqa
where $P(X,V,t)$ is the probability density of the position $X$ and velocity $V$ of the wall as a Brownian piston, and $U(X)$ represents the trapping potential energy. The heat absorption by the 
wall is modeled by the coupling to a Langevin bath \cite{LNP} at the temperature $T_b$ with the coupling, i.e. friction, constant $\gamma_b.$
The collision of the gas particles is represented by the velocity transition rate, $W(V'|V),$ given by 
\[W(V'|V)dV'dt= H(v^\perp-V)\times \rho A(v^\perp-V)dt
\sqrt{\frac{m}{2\pi \kT}}e^{-\frac{m}{2\kT}{v^\perp}^2}\inRbracket{\frac{m+M}{2m}}dV' \]
where $A$ is the surface area of the wall,  $v^\perp$ is the normal component of the incoming velocity of gas particle, and $H(z)$ is the Heaviside unit step function. $v^\perp$ is the function of wall's velocities before ($V$) and after ($V'$) the collision, respectively,  through the momentum conservation rule,
\[V'=V+\frac{2m}{m+M}(v^\perp-V).\]
The truncated equations for the first two moments of $V$ can be derived from the above Master-Boltzmann equation, and the results read
\[ M\frac{d\la V\ra}{dt}=-\la U'(X)\ra -(\gamma+\gamma_b)\la V\ra
+\inRbracket{p-c\frac{\bm{j}^{(e)}}{v_{\rm th}}}A\]
\[ M\frac{d\la V^2\ra}{dt}=-\la VU'(X)\ra -\frac{\gamma}{M}[\la V^2\ra-\kT]
-\frac{\gamma_b}{M}[\la V^2\ra-\kT_b] + c'\la V\ra \]
where $c=\sqrt{\pi/8}$ and the other constant $c'$ is irrelevant as we shall see immediately below.
In the steady state, not only $\la V\ra=d/dt=0$ but 
also $\la VU'(X)\ra$  vanishes. Then the second moment equation tells that 
the kinetic temperature of the Brownian piston, $T_{\rm kin}\equiv M\la V^2\ra,$ is 
given by the well known formula of Langevin dynamics,
\[\kT_{\rm kin}=\frac{\gamma \kT+\gamma_b \kT_b}{\gamma+\gamma_b}\]
Moreover, the second and the third terms on the r.h.s. of the second moment equation
gives the energy transfer to [from] the wall, respectively:
\[j^{(e)}= -\frac{\gamma}{M}[\la V^2\ra-\kT]= \frac{\gamma_b}{M}[\la V^2\ra-\kT_b].\]
With $j^{(e)}$ thus known, the first moment equation in the steady state is nothing but the momentum balance condition:
\[-\la U'(X)\ra +\inRbracket{p-c\frac{j^{(e)}}{v_{\rm th}}}A=0.\]
Our physical reasoning, therefore, reproduces completely the traditional result except for the numerical factor $c$. Moreover, our explanation allows to treat the adiabatic piston, Brownian ratchet models \cite{VdB-prl2004-ratchet}, or inelastic piston \cite{0295-5075-82-5-50008} on the same footing \cite{AFRKKS2012}.

\section{Momentum transfer of a gas with heat transport}
\subsection{Preliminary argument}

The mean free path $\ell_{\rm map}$ of an ideal gas is infinite because the particles undergo no collisions.  Knudsen number $Kn\equiv \ell_{\rm map}/L_{\rm sys}$ is therefore infinite with any system size, $L_{\rm sys}.$ The macroscopic thermo-hydrodynamics \cite{landau-hyd} supposes the opposite limit, $Kn\ll 1.$ 
When we study the thermo-hydrodynamics with energy-transferring boundaries, the physical ideas obtained in the previous section should, therefore, be applicable only to the vicinities of those walls probably with some modifications. The main question is how to reconcile the formula Eq.~\ref{eq:MDD} for the ideal gas with the macroscopic description of thermo-hydrodynamics with non-equilibrium boundary condition.
In this paper we limit ourselves to the steady states with vanishing macroscopic velocity of the gas.
The conservation laws of mass, momentum and energy then impose the constancy of those fluxes.
Below we study first by a purely mechanical toy model that shows the basic compatibility between these fluxes and their nature of symmetry in space and in time (\S\S~\ref{sec:toy}).
Then we go onto the dense hard-core gas with $Kn\ll 1$ (\S\S~\ref{sec:hydro}).

\subsection{Toy model\label{sec:toy}}
\begin{figure}[h] \centerline{\includegraphics[width=2.5cm]{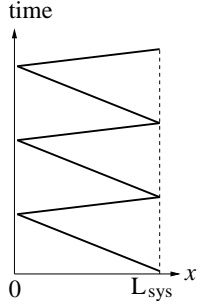}}
\caption{ Space ($x$)- time trajectory of a particle between the hot ($x=0$) and cold ($x=L_{\rm sys}$) walls.} \label{fig:1Dgas}
\end{figure}

We begin by a very elementary kinetic model to discuss the interplay of the energy and momentum transfer\footnote{This is a simplified version of Knudssen heat transfer, see, for example, \cite{gas-Struchtrup2005}, page 25.}.
We take up a single gas particle on the $x$-axis bounded by 
the energy-transferring walls at $x=0$ and at $x={L_{\rm sys}},$ which are macroscopically fixed in space, see Fig.~\ref{fig:1Dgas}.
We further simplify that the hot wall (on the left) receives the particle of velocity $-v_-$  and returns with the velocity $v_+$ with $0<v_-<v_+$. 
The cold wall (on the right) does the opposite operation.
The microscopic mechanism underlying these reflections are irrelevant for our argument. (On might imagine the two tennis player engaging in a rally.)

Before counting the momentum and energy flux, $j^{(e)}$ and $j^{(p)}$, we impose the vanishing of the mass flux,  $j^{(m)}$ in the steady state:
\[ j^{(m)}=(\rho_+ mv_+ - \rho_- mv_-)\hat{\bm{x}}=0.\]
This is satisfied by the densities of rightward and leftward particle,   respectively, 
\[\rho_\pm=\frac{v_\mp}{(v_++v_-){L_{\rm sys}}}.\]
Also the collision frequency on each wall, $\nu_{\rm col}$ is found to be
\[\nu_{\rm col}=\frac{v_+ v_-}{(v_+ +v_-)L_{\rm sys}}.\]
With this $\nu_{\rm col},$ the energy transfer rate $j^{(e)}$ is 
\[ j^{(e)}=\frac{m}{2}(v_+^2-v_-^2)\frac{v_+ v_-}{(v_+ +v_-){L_{\rm sys}}}\hat{\bm{x}}
=\frac{m}{2}(v_+-v_-)\frac{v_+ v_-}{L_{\rm sys}}\hat{\bm{x}} \]
while the momentum flux $j^{(p)}$ reads 
\[ j^{(p)}=m(v_++v_-)\frac{v_+ v_-}{(v_+ +v_-){L_{\rm sys}}}\hat{\bm{x}}\hat{\bm{x}}
=m\frac{v_+ v_-}{L_{\rm sys}}\hat{\bm{x}}\hat{\bm{x}}. \]
We verify that $j^{(e)}$ is odd under time or space-inversion,  while $j^{(p)}$ is even under these operation. The symmetry of $j^{(p)}$ can also be seen from Fig~\ref{fig:1Dgas}.
This model, although simple, shows how the directed energy transfer is established without gradient of momentum flux. In other words, the pressure on the hot and cold walls are the same.

To see more in detail the process {\it at} the walls, we refer to the 
contact value theorem, $p=\kT \rho(0)$   \cite{Henderson1979315},  which give the
equilibrium momentum transfer to a hard wall by a hard core gas in terms of the equilibrium temperature $T$ and the gas particle density at the closest contact surface of the hard wall.
Our interest is the case with energy-transferring walls, see Fig.~\ref{fig:neq-wall.eps}.
When the walls transfer the energy, the $p$ should indicate the total momentum flux 
$j^{(p)}$, i.e.,
\[p_{\rm neq}=|j^{(p)}|=m ({{v_+}}+{{v_-}})\nu_{\rm col}\]
 and $\kT_{\rm neq}/2$ the kinetic energy per particle, 
\[ \frac{\kT_{\rm neq}}{2}= \frac{m}{2}\,
\frac{\rho_+ {v_+}^2 +\rho_-  {v_-}^2}{\rho_+ +\rho_-}
= \frac{m}{2}v_+ v_-.\]

Since the total density $\rho(0)$ on the wall is, by the homogeneity,
\[\rho(0)=\inRbracket{\inv{{v_+}}+\inv{{v_-}}}\nu_{\rm col}=\inv{L_{\rm sys}}\]
we arrive at a form of the contact value theorem in non-equilibrium.
\eqn{\label{eq:neq-contact}p_{\rm neq}=\kT_{\rm neq}\,\rho(0).}
In other words, while the symmetry allows the correction to the r.h.s. of the form $\propto (
v_+-v_-)^2$ or $\propto {j^{(e)}}^2$, the contact value theorem holds up to the order of 
${\cal O}({j^{(e)}}^2)$ if $p_{\rm eq}$ and $T_{\rm eq}$ are appropriately chosen (cf.\cite{KNST-prl2008}).

\subsection{Non-equilibrium hydrodynamics \label{sec:hydro}}
In the non-equilibrium steady state with heat flux of a dense hard-core gas with $Kn\ll 1$, the energy flux vector field, $\bm{j}^{(e)}$, and momentum flux tensor field, $ {\sf j}^{(p)}$,  
 must satisfy the basic conservation laws:
\[\nabla\cdot\bm{j}^{(e)}=0,\qquad \nabla\cdot {\sf j}^{(p)}={\bf 0}.\]
If the wall is perpendicular to the $x$-axis, the system is homogeneous in $y$ and $z$ directions and the above conditions are reduces to
\[ j^{(e)}_x=\mbox{const.}\qquad {\sf j}^{(p)}_{xx}=\mbox{const.}\]
 If the heat conduction obeys approximately the Fourier's law, $\bm{j}^{(e)}=k_{T}\nabla T,$ with 
 $k_T$ being the heat conductivity, the temperature gradient keeps constancy of the energy flux.
 As for the momentum flux ${\sf j}^{(p)},$
  the symmetry argument or Curie principle \cite{curie} 
 allows the anisotropy of the type ${\sf j}^{(p)}=p{\bf 1}+a (\hat{\bm{x}}\hat{\bm{x}}-
 \inv{3}{\bm 1})$ with  $a$ characterizing the deviatoric part of the flux due to heat flux $\| \bm{x}$.
 However, seeing that the ${\cal O}({j^{(e)}}^2)$ contribution was missing in the above simple model Eq.~\ref{eq:neq-contact}, we simply identify  ${\sf j}^{(p)}=p_{\rm neq}{\bf 1}$ to be the pressure in the present approximation.
If $p_{\rm neq}$ obeys approximately the equilibrium equation of state, $p= p(\rho,T),$ among the pressure $p$, temperature $T$ and the density $\rho$, the density $\rho(\bm{x})$ varies in a manner locally compensating the heterogeneity of the temperature $T(\bm{x})$ so that the $p_{\rm neq}$ remains homogeneous.

Our concern is how we can relate the momentum flux and the energy flux in the dense hard-core gas where $p_{\rm neq}$ reflects already both the incoming and outgoing particles. 
Below we will indicate that the relation like Eq.~\ref{eq:MDD} corresponds to the skewness of the velocity distribution of particles (especially) at the energy-transferring wall.
To be concrete we imagine the dense hard-core gas which is conducting the heat rightwards up to the energy-transferring wall at $x=0$ without convection (Fig.~\ref{fig:neq-wall.eps}). We also assume that the wall exchanges only
the $x$-component of momentum.
Now we introduce the velocity distribution function $f(v_x;\bm{x})$ per unit volume of gas particles. 
Then the particle density $\rho(\bm{x})$ is given by 
\[\rho(\bm{x}) =\int f(v_x;\bm{x}) dv_x.\]
The conditions on the fluxes of mass, momentum and energy along the $x$ axis are given,
respectively, as
\beqa 
0&=&\bm{j}^{(m)}(\bm{x})\cdot\hat{\bm{x}}  = \int m v_x f(v_x;\bm{x})dv_x 
\cr
p_{\rm neq} &=&\hat{\bm{x}}\cdot {\sf j}^{(p)}(\bm{x})\cdot \hat{\bm{x}}= \int m v_x^2 f(v_x;\bm{x})dv_x 
\cr 
j^{(e)}&=& \bm{j}^{(e)}(\bm{x})\cdot \hat{\bm{x}}=    \int \frac{m}{2} v_x^3 f(v_x;\bm{x})dv_x, 
\eeqa
where $p_{\rm neq}$ and $j^{(e)}$ are independent of the position $\bm{x}$.

Now we focus on the thin slab of the distance $\ll \ell_{\rm mfp}$ from the energy-transferring wall. In this slab we assume that the gas particles undergo practically no collisions except for with the wall.
We introduce the partial momentum fluxes associated to the {\it incoming} particles, $j^{(p)}_{\rm in}$ and to the {\it outgoing} particles, $j^{(p)}_{\rm out},$  right before the wall:
\[\left.j^{(p)}_{\rm in}\right|_{x=0-} \equiv \int mv_x^2 H(+v_x)f(v_x;\bm{x}) dv_x,
\qquad
\left.j^{(p)}_{\rm out}\right|_{x=0-} \equiv \int mv_x^2 H(- v_x)f(v_x;\bm{x}) dv_x.
\]
In equilibrium where $f(v_x;\bm{x})$ are symmetric with respect to $v_x$, the both partial fluxes are the same. In the presence of the heat flux it is no more the case.
While the asymmetric velocity distribution for Knudsen heat transfer, i.e. the above toy model, is usually singular and far from Maxwellian, the collisions make the velocity distribution looks like 
skewed Maxwell distribution. We, therefore, assume an approximate form\footnote{On the wall, $x=0$, the very MDD implies the discontinuity in $f(v_x;\bm{x})$ at $v_x=0$. Here, however, we shall use a smoothed form as qualitative model. See also, for example, \cite{gas-Struchtrup2005}, page 202.}: 
\[ \left.f(v_x;\bm{x})\right|_{x=0-}=\inSbracket{
c_0+ c_1 \inRbracket{\frac{v_x}{\sigma}}
+ c_2 \inRbracket{\frac{v_x}{\sigma}}^2+ c_3 \inRbracket{\frac{v_x}{\sigma}}^3}
\,e^{-\frac{{v_x}^2}{2\sigma^2}}.\]
For a week energy flux, the terms containing $c_1$, $c_2$ and $c_3$ are regarded to be small perturbations with  respect to the main term $c_0$.
The expression of the density and the flux conditions mentioned above impose
\beqa
\frac{\left.\rho\right|_{x=0-}}{\sqrt{2\pi}}&=& (c_0 +c_2)\sigma 
\cr
0&=& c_1+3c_3 
\cr \frac{p_{\rm neq}}{\sqrt{2\pi}m}&=&(c_0 +3c_2) \sigma^3
\cr \frac{2j^{(e)}}{3\sqrt{2\pi}m}&=&(c_1+ 5c_3) \sigma^5=2c_3 \sigma^5,
\eeqa
where in the last line we used the vanishing mass flux condition in the second line; $c_1=-3c_3.$
    Finally the difference between the partial momentum fluxes $j^{(p)}_{\rm in}$ and $j^{(p)}_{\rm out}$ read
\[\frac{(j^{(p)}_{\rm in}-j^{(p)}_{\rm out})|_{x=0-}}{2m}=(c_1+4c_3)\sigma^4=c_3\sigma^4,\]
where again we used the vanishing mass flux condition.
    Then if we introduce the squared average of the particle velocity (noting $|c_2|/c_0\ll 1$ for weak non-equilibrium),
\[\overline{v_x^2}|_{x=0-}=\frac{p_{\rm neq}}{\left.\rho\right|_{x=0-}}=\frac{1+3c_2/c_0}{1+c_2/c_0}\sigma^2\simeq \sigma^2,\]
we obtain the relation of MDD and the energy flux, reminiscent of Eq.~\ref{eq:MDD}:
\[\left. \inRbracket{{j^{(p)}_{\rm in}-j^{(p)}_{\rm out}}}\right|_{x=0-}
=C\frac{j^{(e)}}{\inSbracket{\overline{v_x^2}|_{x=0-}}^{1/2}}\]
with a numerical factor $C=\sqrt{2}/(3\!\sqrt{\pi}),$ which is subject to our approximations.

  In onclusion, the dense hard-core gas conducting the heat carries also momentum through the asymmetric distribution of particle's velocity, and the concept of MDD is a neat way to explain the relation between the energy flux and the partial momentum fluxes in the non-equilibrium steady state.\\

We would like to acknowledge the organizers of the 25th Smoluchowski Symposium.
\bibliographystyle{apsrev.bst}    
\bibliography{bib-AF}
\end{document}